\documentclass[11pt]{article}
\usepackage{amssymb,amsfonts,amsmath,latexsym,graphicx}

\newtheorem{theorem}{Theorem}
\newtheorem{definition}[theorem]{Definition}
\newtheorem{corollary}[theorem]{Corollary}
\newtheorem{lemma}[theorem]{Lemma}

\newtheorem{proposition}[theorem]{Proposition}

\newcommand{\email}[1]{{\small E-mail: {\textsf {#1}}}}
\newcommand{\http}[1]{{\small Internet: {\textsf {#1}}}}
\newcommand{\affil}[1]{{\small\sl #1}}
\newcommand{\keywords}{\medskip\noindent{\bf Keywords. }}
\newcommand{\AMS}{\noindent{\small\bf AMS classification (2000). }}
\newcommand{\be}{\begin{equation}}
\newcommand{\ee}{\end{equation}}
\newcommand{\N}{\mathbb{N}}

\newcommand{\BbR}{\mathbb{R}}
\def\BbC{\mathbb{C}}
\newcommand{\Z}{\mathbb{Z}}
\def \1{{\rm 1\kern-4pt 1}}

\newcommand{\beq}{\begin{eqnarray}}
\newcommand{\eeq}{\end{eqnarray}}
\newcommand{\bq}{\begin{equation}}
\newcommand{\eq}{\end{equation}}

\newcommand{\Sum}{\displaystyle \sum}

\newcommand{\Int}{\displaystyle \int}
\newcommand{\Frac}{\displaystyle \frac}
\newcommand{\Inf}{\displaystyle \inf}
\newcommand{\Sup}{\displaystyle \sup}
\newcommand{\Lim}{\displaystyle \lim}

\newcommand{\Min}{\displaystyle \min}
\def\un { \rm 1\!\!I}

\begin{document}
\title{\sl Some connections between Dirac-Fock and Electron-Positron Hartree-Fock}
 \author{Jean-Marie BARBAROUX\\
 \affil{CPT-CNRS}\\
 \affil{Luminy Case 907 - 13288 Marseille
 C\'edex 9 - France}\\
 \email{barbarou@cpt.univ-mrs.fr}\\
 Maria J. ESTEBAN, Eric SERE\\
 \affil{Ceremade (UMR CNRS no. 7534), Universit\'e Paris
 IX-Dauphine,}\\
 \affil{Place de Lattre de Tassigny, 75775 Paris
 C\'edex~16, France}\\
 \email{esteban, sere@ceremade.dauphine.fr}\\
 \http{http://www.ceremade.dauphine.fr/\raisebox{-4pt}{
 $\widetilde{\phantom{x}}$}esteban/, \raisebox{-4pt}{
 $\widetilde{\phantom{x}}$}sere/} }
 \date{\today}
 \maketitle
\thispagestyle{empty}
\abstract{\sl  We study the ground state solutions of the
Dirac-Fock model in the case of weak electronic repulsion, using
bifurcation theory. They are solutions of a min-max problem. Then
we investigate a max-min problem coming from the electron-positron
field theory of Bach-Barbaroux-Helffer-Siedentop.
We show that given a radially symmetric nuclear charge, the
ground state of Dirac-Fock solves this max-min problem for certain
numbers of electrons. But we also exhibit a situation in
which the max-min level does not correspond to a solution of the
Dirac-Fock equations together with its associated self-consistent projector.}\smallskip

\medskip

\AMS{\scriptsize Primary:
Secondary:
}


\vspace{6mm}

\noindent {\bf 1 - Introduction}.

\vspace{2mm} The electrons in heavy atoms experience important
relativistic effects. In computational chemistry, the
Dirac-Fock (DF) model \cite{[swi]}, or the more accurate
multiconfiguration Dirac-Fock model \cite{[lin-ro]},  take these effects into account. These models
are built on a multi-particle Hamiltonian which is in principle
not physically meaningful, and whose essential spectrum is the
whole real line. But they seem to function very well in
practice, since approximate bound state solutions are found and
numerical computations are done and yield results in quite good
agreement with experimental data (see e.g. \cite{[gor-al]}).
Rigorous existence results for solutions of the DF equations can
be found in \cite{Esteban-Sere-99} and \cite {pat}. An
important open question is to find a satisfactory physical
justification for the DF model.

\medskip
It is well known that the correct theory including quantum and
relativistic effects is quantum electrodynamics (QED). However,
this theory leads to divergence problems, that are only solved in
perturbative situations. But the QED equations in heavy atoms are
nonperturbative in nature, and attacking them directly seems a
formidable task. Instead, one can try to derive approximate models
from QED, that would be adapted to this case. The hope is to show
that the Dirac-Fock model, or a refined version of it, is one of
them. Several attempts have been made in this direction
 (see \cite{Mittleman-81,[suc1], [suc2],cha-ir} and the
references therein).  Mittleman  \cite{Mittleman-81}, in
particular, derived the DF equations with ``self-consistent
projector" from a variational procedure applied to a QED
Hamiltonian in Fock space, followed by the standard Hartree-Fock
approximation. More precisely, let $H^c$ be the free Dirac
Hamiltonian, and $\Omega$ a perturbation. We denote
$\Lambda^+(\Omega)=\chi_{(0,\infty)}(H^c+\Omega)$. The electronic
space is the range ${\cal H}^+(\Omega)$  of this projector. If one
computes the QED energy of Slater determinants of $N$ wave
functions in this electronic space, one obtains the DF energy
functional restricted to $({\cal H}^+(\Omega))^N$. Let
$\Psi_\Omega$ be a minimizer of the DF energy in the projected
space $({\cal H}^+(\Omega))^N$ under normalization constraints. It
satisfies the projected DF equations, with projector
$\Lambda^+(\Omega)$. Let $E(\Omega):=\mathcal{E}(\Psi_\Omega)$.
Mittleman showed (by formal arguments) that
the stationarity of $E(\Omega)$ with respect
to $\Omega$ implies that $\Lambda^+(\Omega)$
coincides, {\sl on the occupied orbitals}, with the
self-consistent projector associated to the mean-field Hartree-Fock Hamiltonian created by $\Psi_\Omega$.  From this he infers (\cite{Mittleman-81}, page 1171) : {\sl ``Hence, $\Omega\,$ is the Hartree-Fock
potential when the Hartree-Fock approximation is made for the wave function".}
 \medskip

Recently  rigorous mathematical results have been obtained in a
series of papers by Bach {\it et al.} and Barbaroux {\it et al.}
\cite{Bach-Barbaroux-Helffer-Siedentop-98,
Bach-Barbaroux-Helffer-Siedentop-99,
Barbaroux-Farkas-Helffer-Siedentop-XX}  on a Hartree-Fock type
model involving electrons and positrons. This model (that we will
call EP) is related to the works of Chaix-Iracane \cite{cha-ir}
and Chaix-Iracane-Lions
 \cite{cha-ir-li}. Note, however, that in \cite{Bach-Barbaroux-Helffer-Siedentop-98,
Bach-Barbaroux-Helffer-Siedentop-99,
Barbaroux-Farkas-Helffer-Siedentop-XX} the vacuum polarisation is
neglected, contrary to the Chaix-Iracane approach. In
\cite{Bach-Barbaroux-Helffer-Siedentop-98}, in the case of the
vacuum, a max-min procedure in the spirit of Mittelman's work is
introduced.  In \cite{Barbaroux-Farkas-Helffer-Siedentop-XX}, in
the case of $N$-electron atoms, it is shown that critical pairs
$(\gamma, P^+)$ of the electron-positron Hartree-Fock energy
${\cal E}_{EP}$ give solutions of the self-consistent DF equations. This result is
an important step towards a rigorous justification of Mittleman's
ideas. All this suggests, in the case of $N$-electrons atoms, to
maximize the minimum $E(\Omega)$ with respect to $\Omega$. It is
natural to expect that this max-min procedure gives solutions of
the DF equations, the maximizing projector being the positive projector of the self-consistent
Hartree-Fock Hamiltonian.  We call this belief (expressed here in rather
imprecise terms) ``Conjecture M".\medskip

In \cite{Esteban-Sere-01} and \cite{Esteban-Sere-02}, when
analyzing the nonrelativistic limit of the DF equations, Esteban
and S\'er\'e derived various equivalent variational problems
having as solution an ``electronic" ground state for the DF
equations. Among them, one can find min-max and max-min
principles. But these principles are nonlinear, and do not solve
Conjecture M.\medskip

In this paper we try to give a precise formulation of  Conjecture M in the spirit of Mittleman's
ideas and to see if it holds true or not, in the limit case of small interactions between
electrons.  We prove that in this perturbative regime, given a radially symmetric nuclear potential, Conjecture M may hold or not depending on the number of electrons. The type of ions which are covered by our study are those  in which the number of electrons is much smaller than the number of protons in the nucleus, with, additionally, $\,c$ (the speed of light) very large.

\medskip
The paper is organized as follows : in \S 2 we introduce the notations and state our main results (Theorems \ref{[8]} and \ref{TT}). Sections 3 and 4 contain the detailed proofs.

\vspace{6mm}

\noindent {\bf 2 - Notations and main results}.

\vspace{2mm}

\vspace{1mm} In the  whole paper we choose a system of units in which Planck's constant, $\,\hbar\,$,
and the mass of the electron are equal to $1$ and $\,Ze^2=4\pi\epsilon_0$, where $Z$ is the number of protons in the nucleus. In this system of units, the Dirac
Hamiltonian can be written as
\begin{equation}
   H^c \ = \ -ic\,{\boldsymbol{\alpha}}
   \cdot \nabla + c^2 \beta, \label{(1)}
\end{equation}
where $c>0$ is the speed of light , $\beta = \left (
\begin{array}{cc}
\1 & 0 \\ 0 & -{\1} \\
\end{array} \right)$,
$\boldsymbol{\alpha}=(\alpha_1, \alpha_2,\alpha_3)$, ${\alpha_\ell} = \left
( \begin{array}{cc} 0 & \sigma_\ell
\\
\sigma_\ell & 0 \\
\end{array} \right)$  and   the  $\sigma_\ell $'s  are the
Pauli matrices. The operator $H^c$ acts on 4-spinors, i.e.
functions from $\BbR^3$ to $\BbC^4$, and it is self-adjoint in
$L^2(\BbR^3,\BbC^4)$, with domain $H^1(\BbR^3,\BbC^4)$ and
form-domain $H^{1/2}(\BbR^3,\BbC^4)$. Its spectrum is the set
$(-\infty,-c^2]\cup [c^2,+\infty)$.\medskip

In this paper, the charge density of the nucleus  will be  a smooth, radial and compactly supported
nonnegative function $n$,  with
$\int n = 1$, since in our system of units $Ze^2=4\pi\epsilon_0$. The corresponding Coulomb potential is
$V:=-n*(1/|x|)$. Then $V : \BbR^3 \rightarrow (-\infty, 0)$ is a
smooth negative radially symmetric potential such that
 $$
  - \Frac{1}{|x|} \leq V(x) < 0 \quad (\forall x)\quad,
  \quad |x|\, V(x) \simeq -1\;\hbox{ for} \quad |x|\;\hbox{ large
  enough}\;.
 $$
Note that the smoothness condition on $V$ is only used in step 3
of the proof of Proposition~\ref{conj}. Actually we believe that
this condition can be removed.

\medskip
 It is well known that $H^c+V$ is essentially self-adjoint
and for $c>1$, the spectrum of this operator is as follows:
 $$
  \sigma (H^c + V )= (- \infty, - c^2]
  \cup \{\lambda_1^c, \lambda_2^c, \dots \}
  \cup [c^2, + \infty),
 $$
with $0< \lambda_1^c < \lambda_2^c < \dots$ and $\Lim_{\ell
\rightarrow + \infty} \lambda_\ell^c = c^2$.

Finally define the spectral subspaces ${\cal M}^c_i = \mathrm{Ker}
(H^c + V - \lambda_i^c \,\1)$ and let $N_i^c$ denote ${\cal
M}^c_i$'s dimension.

Since the potential is radial, it is well known that the
eigenvalues $\lambda^c _i$ are degenerate (see {\it e.g.}
\cite{Thaller-92}). For completeness, let us explain this in some
detail. To any $A\in SU(2)$ is associated a unique rotation $R_A
\in SO(3)$ such that $\forall x\in\BbR^3$,
$(R_Ax)\cdot\sigma=A(x\cdot \sigma)A^{-1}$, where
$\sigma=(\sigma_1, \sigma_2,\sigma_3)$. This map is a morphism of
Lie groups. It is onto, and its kernel is $\{I,-I\}$. It leads to
a natural unitary representation $\bullet$ of $SU(2)$ in the
Hilbert spaces of 2-spinors $L^2(S^2,{\BbC}^2)$ and
$L^2(\BbR^3,{\BbC}^2)$, given by
\begin{equation} \label{rep2}
   (A\bullet \phi) (x):=A\,\phi(R_A^{-1}x)\;.
\end{equation}
Then, on the space of 4-spinors
$L^2(\BbR^3,{\BbC}^4)=L^2(\BbR^3,{\BbC}^2)\oplus
L^2(\BbR^3,{\BbC}^2)$, one can define the following unitary
representation (denoted again by $\bullet$)
\begin{equation} \label{rep4} \
   \Big (A\bullet {{ \phi}\choose {\chi
   }}\Big) (x):= {{ (A\bullet\phi)(x)}\choose {(A\bullet \chi)(x) }}=
   {{ A\phi(R_A^{-1}x)}\choose {A\chi(R_A^{-1}x) }}\;.
\end{equation}
The radial symmetry of $V$ implies that $H^c+V$ commutes with
$\bullet$. The eigenspaces ${\cal M}^c_i$ are thus $SU(2)$
invariant. Now, let $\hat{J}=(\hat{J}_1,\hat{J}_2,\hat{J}_3)$ be
the total angular momentum operator associated to the
representation $\bullet$. The eigenvalues of
$\hat{J}^2=\hat{J}_1^2+\hat{J}_2^2+\hat{J}_3^2$ are the numbers
$(j^2-1/4)$ , where $j$ takes all positive integer values. If
$\phi$ is an eigenvector of $\hat{J}^2$ with eigenvalue
$(j^2-1/4)$ , then the $SU(2)$ orbit of $\phi$ generates an
$SU(2)$ invariant complex subspace of dimension $2j\geq 2$. This
implies the following fact, which will be used repeatedly in the
present paper:

\begin{lemma}
If $\phi\in L^2(\BbR^3,{\BbC}^2)$ is not the zero function, then
there is $A\in SU(2)$ such that $\phi$ and $A\bullet \phi$ are two
linearly independent functions.
\end{lemma}

\noindent {\bf Proof of the Lemma.} Assume, by contradiction, that
${\BbC}\,\phi\,$ is $SU(2)$ invariant. Then $\phi$ is an
eigenvector of $J_\ell$ for $\,\ell=1,2,3$, hence it is eigenvector of
$\hat{J}^2$. But we have seen that in such a case, the $SU(2)$
orbit of $\phi$ must contain at least two independent vectors:
this is absurd. $\Box$
\medskip

As a consequence of the Lemma, the spaces ${\cal M}^c_i$ have
complex dimension at least 2. The degeneracy is higher in general:
for each $j\geq 1$ , $H^c+V$ has infinitely many eigenvalues of
multiplicity at least $2j$. Note that in the case of the Coulomb
potential, the eigenvalues are even more degenerate (see  {\it
e.g.} \cite{Thaller-92}).
\medskip
Now, on the Grassmannian manifold
 $$
   G_N (H^{1/2}) := \lbrace W \
   {\rm subspace \ of \ } \ H^{1/2} (\BbR^3, \BbC^4);
   \dim_{\BbC} \,(W) = N \rbrace
 $$
we define the Dirac-Fock energy ${\cal E}_{\kappa}^c$ as follows
\begin{eqnarray}
 {\cal E}_{\kappa}^c (W) \!\!\!\!\!& : = {\cal E}_{\kappa}^c
(\Psi) : =
\Sum^N_{i = 1}
\Int_{\BbR^3} ( (H^c + V) \psi_i, \psi_i) dx + \nonumber \\
 \label{1} \\
\qquad+ &\!\!\!\! \Frac{{\kappa}}{2} \iint_{\BbR^3 \times \BbR^3}
\Frac{ \rho_{\Psi} (x) \rho_{\Psi} (y) -
|R_{\Psi}(x,y)|^2}{|x-y|} dx dy \ ,  \nonumber\\
 \nonumber \end{eqnarray}
where $\kappa>0$ is a small constant, equal to $\,e^2/4\pi\epsilon_0\,$ in our system of units,  $\{ \psi_1, \dots \psi_N\}$
is any orthonormal basis of $W$, $\Psi$ denotes the $N$-uple
$(\psi_1, \dots \psi_N)$,  $\rho_\Psi$ is a scalar and $R_\Psi$ is
a $4 \times 4$  complex matrix, given by
\begin{equation}
  \rho_{\Psi}(x) \ = \ \displaystyle{\sum^N_{\ell = 1}} \ \Bigl( \psi_{_\ell} (x),
  \psi_{_\ell} (x) \Bigr)\;,\quad
  R_{\Psi}(x,y) \ = \ \displaystyle{\sum^N_{\ell = 1 }} \ \psi_{_\ell} (x) \otimes
  \psi^\ast_{_\ell} (y) \, .
\end{equation}

Saying that the basis $\{ \psi_1, \dots \psi_N\}$ is orthonormal
is equivalent to saying that
\begin{equation} \label{2}
   {\rm Gram}_{L^2} \Psi = \1_N \;.
\end{equation}

We will use interchangeably the notations ${\cal E}_{\kappa}^c (W)$
or ${\cal E}_{\kappa}^c (\Psi)$. The energy can be considered as a
function of $W$ only, because if $u\in U(N)$ is a unitary matrix,
\begin{equation} \label{3}
   {\cal E}_{\kappa}^c (u \Psi) = {\cal E}_{\kappa}^c
   (\Psi) \ .
\end{equation}
with the notation $(u\Psi)_k=\sum_l u_{kl}\psi_l$.

Note that since $V$ is radial, the DF functional is also invariant
under the representation $\bullet$ defined above. Its set of
critical points will thus be a union of $SU(2)$ orbits.
\medskip

Finally let us introduce a set of projectors as follows:

\begin{definition} \label{P}
Let $P$ be an orthogonal projector in $L^2 ( \BbR^3, \BbC^4 )$,
whose restriction to $H^{\frac{1}{2}} ( \BbR^3, \BbC^4 )$ is a
bounded operator on $H^{\frac{1}{2}} ( \BbR^3,\BbC^4 )$. Given
$\varepsilon> 0$, $P$ is said to be $\varepsilon$-close  to
$\Lambda_+^c:=\chi_{(0,+\infty)}(H^c)\,$ if and only if, for all
$\psi \in H^{\frac{1}{2}} (\BbR^3, \BbC^4)$, $$ \Big \Vert
\Bigl(-c^2 \Delta + c^4 \Bigr)^{\frac{1}{4}} \Bigl(P- \Lambda^c_+
\Bigr) \psi \Big \Vert _{L^2 (\BbR^3, \BbC^4)}
    \leq  \varepsilon \  \Big \Vert \Bigl(-c^2 \Delta + c^4
\Bigr)^{\frac{1}{4}}  \psi \Big \Vert _{L^2 (\BbR^3, \BbC^4)} \ .$$
\end{definition}

In  \cite{Esteban-Sere-01} the following result is proved :

\begin{theorem}[\cite{Esteban-Sere-01}]\label{th:ES1}
Take $V\;,\;N$ fixed. For $c$ large and $\epsilon_0, \ {\kappa}$
small enough, for all $\,P\,$ $\,\varepsilon_0$-close to $\,\Lambda^c_+$,
$$ c(P) := \Inf_{W^+ \in G_N (P H^{1/2})} \sup_{W \in G_N (H^{1/2}) \atop P (W) =
W_+}  {\cal E}_{\kappa}^c(W)$$
is independent of $P$ and we denote it by $\,E^c_{\kappa} $. Moreover,
 $E^c_{\kappa}$ is achieved by a solution $W_{\kappa}=$span$\{ \psi_1, \dots \psi_N \}$
of the Dirac-Fock equations:
\renewcommand{\theequation}{DF}
\setcounter{equation}{0}
\begin{equation} \label{DF}
    \left\lbrace
     \begin{array}{ll}
        H^c_{{\kappa}, W_{\kappa}} \psi_i = \epsilon^c_i
        \,\psi^c_i , \ 0 < \epsilon^c_i < 1 , \\
        {\rm Gram}_{L^2} \Psi = 1_{\N}\\
\end{array} \right.
\end{equation}
with
\renewcommand{\theequation}{MF}
\setcounter{equation}{0}
\begin{equation} \label{MF}
    H^c_{{\kappa}, W}\, \varphi := (H^c + V +
    {\kappa} \,\rho_{\Psi} \ast \Frac{1}{|x|}) \varphi - {\kappa}
   \Int_{\BbR^3} \Frac{R_\Psi (x,y) \varphi (y)}{|x-y|} dy\; .
\end{equation}
\end{theorem}

\medskip\noindent{\bf Remark.}
It is easy to verify that $\,\varepsilon_0>0\,$  given,
for $\,c$ large and $\kappa\,$ small enough, $\chi_{(0,\infty)}
(H^c_{{\kappa}, W_{\kappa}})$ is $\,\varepsilon_0$-close to $\,\Lambda^+_c$.
\begin{corollary}[\cite{Esteban-Sere-01}]\label{[2]}
Take $V,\, N$ fixed. Choose $c$ large and ${\kappa}$ small enough.
If we define the projector
 $$
  P^+_{\kappa,W} = \chi_{(0,\infty)} (H^c_{{\kappa}, W})
 $$
with $ H^c_{{\kappa}, W}$ given by formula (\ref{MF}),
then
\begin{equation} \label{4}
    E^c_{\kappa} = \Min_{W \in G_n (H^{1/2})
    \atop P^+_{\kappa,W} W = W} {\cal E}^c_{\kappa}
    (W) = \Min_{{W \in G_N (H^{1/2})} \atop W \ {\rm
    solution \ of  \ (DF)}} {\cal E}_{\kappa}^c ( W) \ .
\end{equation}

\end{corollary}

\medskip

Another variational problem was introduced in the works of Bach
{\it et al.} and Barbaroux {\it et al.}
(\cite{Bach-Barbaroux-Helffer-Siedentop-98,Bach-Barbaroux-Helffer-Siedentop-99,Barbaroux-Farkas-Helffer-Siedentop-XX})
: define

\begin{equation}
  \widetilde{\cal P}_\kappa = \{ P^+_{\kappa,\widetilde{W}} =
  \chi_{[0, \infty)} (H^c_{{\kappa}, \widetilde W}) \; ;\;\widetilde{W}\in G_N(H^{1/2})\},
\end{equation}
and

\begin{eqnarray*}
  S^{N}_{\kappa,\widetilde{W}} & := & \{\gamma\in {S}_1 (L^2)\, ,\
  \gamma=\gamma^*\, , \ H^c_{{\kappa}, \widetilde{W}}\gamma \in S_1\, , \\
  & & \ \  P^+_{\kappa,\widetilde{W}} \gamma P^-_{\kappa,\widetilde{W}}=0\,,\
  -P^-_{\kappa,\widetilde{W}}  \leq \gamma \leq
  P_{\kappa,\widetilde{W}}^+ \, ,\ {\rm tr}\,\gamma = N \},
\end{eqnarray*}
with the notation
$P^-_{\kappa,\widetilde{W}}:=\un-P^+_{\kappa,\widetilde{W}}$, and $S_1$
being the Banach space of trace-class operators on
$L^2(\BbR^3,{\BbC}^4)$. For all $\gamma\in
S^{N}_{\kappa,\widetilde{W}}$, let
 $$
  {\cal F}^c_{\kappa}(\gamma) = {\rm tr}\left((H^c + V )\gamma\right) +
  \frac{{\kappa}}{2} \int \frac{\rho_\gamma({x})\rho_\gamma({
  y})}{| {x} - { y}|}{\rm d}{ x}\, {\rm d}{y} -
  \frac{{\kappa}}{2} \int \frac{|\gamma(x,y)|^2}{| { x} - { y}|}{\rm
  d}x\, {\rm d} y.
 $$
Here,  $\rho_\gamma({ x}):= \sum_{s=1}^4\gamma_{s,s}(x,\,x) =
\sum_n w_n |\psi_n({ x})|^2$, with $w_n$ the eigenvalues of
$\gamma$ and $\psi_n$ the eigenspinors of $\gamma$, and
$\gamma(x,y) = \sum_n w_n \psi_n(x)\otimes \overline{\psi_n}(y)$,
i.e., $\gamma(x,y)$ is the kernel of $\gamma$.

\medskip
In \cite{Barbaroux-Farkas-Helffer-Siedentop-XX}  it has been
proved that for every $\,P^+_{\kappa,\widetilde{W}}\in \widetilde{\cal
P}_\kappa \,$, the infimum of $\,{\cal F}_\kappa^c\,$ on the set
$\, S^{N}_{\kappa,\widetilde{W}}\,$  is actually equal to the infimum
defined in the smaller class of Slater determinants. More
precisely, with the above notations,

\begin{theorem}[\cite{Barbaroux-Farkas-Helffer-Siedentop-XX}]\label{TTT}
For ${\kappa}$ small enough and for
all $P^+_{\kappa,\widetilde{W}} \in \widetilde{\cal P}_\kappa$, one has
\begin{equation}
   \inf_{\gamma\in S^{N}_{\kappa,\widetilde{W}}}  {\cal F}^c_{\kappa}(\gamma) =
   \inf_{W\in G_N(P^+_{\kappa,\widetilde{W}} H^{1/2})} {\cal E}^c_{\kappa} (W)
\end{equation}
Moreover, the infimum  is achieved by a solution of the projected
Dirac-Fock equations, namely
 $$
  \gamma_{min} = \sum_{i=1}^N \langle\psi_i\, , .\rangle \psi_i
 $$ with $P^+_{\kappa,\widetilde{W}}\psi_i =\psi_i$ ($i=1,\ldots N)$, and for
$W_{\min} := span(\psi_1,\ldots, \psi_N)$ ,
\begin{equation}
  \left\lbrace \begin{array}{ll} P^+_{\kappa,\widetilde{W}} H^c_{{\kappa}, W_{\min}}
  P^+_{\kappa,\widetilde{W}} \psi_i = \epsilon_i \psi_i , \ 0 < \epsilon_i < 1 , \\
  {\rm Gram}_{L^2} \Psi = 1_{\N}\\
  \end{array} \right.
\end{equation}
\end{theorem}

\medskip

Let us now define the following sup-inf:
\begin{equation}\label{supinf}
  e^c_{\kappa} : =
  \Sup_{P^+_{\kappa,\widetilde{W}} \in \widetilde{\cal P} }
  \quad \inf_{W\in G_N(P^+_{\kappa,\widetilde{W}} H^{1/2})} {\cal E}^c_{\kappa} (W)\,.
\end{equation}

Then, Theorem \ref{TTT}
has the following consequence:
\begin{corollary}
If ${\kappa}$ is small enough,
 $$
   e^c_{\kappa} = \sup_{P^+_{\kappa,\widetilde{W}}
   \in \widetilde{\cal P}_\kappa} \;\;\inf_{\gamma\in
   S^{N}_{\kappa,\widetilde{W}}} {\cal F}^c_{\kappa}(\gamma)
 $$
\end{corollary}

From the above definitions, Theorem~\ref{th:ES1},
Corollary~\ref{[2]} and the remark made after
Theorem~\ref{th:ES1}, we clearly see that for all ${\kappa}$ small
and $c$ large,
\begin{equation} \label{6}
   E^c_{\kappa} \geq
   e^c_{\kappa}.
\end{equation}
\noindent One can hope more:
\medskip

\noindent {\bf Conjecture M:} {\it The energy levels $\,
E^c_{\kappa}\,$ and $\, e^c_{\kappa}\,$ coincide, and there is a
solution $W^c_{\kappa}$ of the DF equations such that $${\cal
E}^c_\kappa(W^c_{\kappa})=e^c_\kappa=\inf_{V \in
G_N(P^+_{\kappa,W^c_{\kappa}} H^{1/2})} {\cal E}^c_{\kappa}
(V)\,.$$ In other words, the max-min level $e^c_\kappa$ is
attained by a pair $(W,P^+_{\kappa,\widetilde{W}})$ such that
$\widetilde{W}=W$.}
\medskip

This paper is devoted to discussing this conjecture, which, if it
were true, would allow us to interpret the Dirac-Fock model as a
variational approximation of QED.
\medskip
In order to study the different cases that can appear when
studying the problems $\,E^c_{\kappa}\,$ and $\,e^c_{\kappa}\,$
for $\,{\kappa}\,$ small, we begin by discussing the case
$\,{\kappa}=0$.

\medskip
\begin{proposition}\label{[5]}
 Conjecture M is true in the case ${\kappa} = 0$.
\end{proposition}

\noindent {\bf Proof}. The case $\kappa=0$ is obvious. Indeed, all
projectors $P^+_{0,\widetilde{W}}$ coincide with the projector
$\chi_{[0,\infty)}(H^c+V)$. The level $E_0^c$, seen as the minimum
of Corollary 2, is achieved by any  $N$-dimensional space
$W_{min}$ spanned by $N$ orthogonal eigenvectors of $H^c+V$ whose
eigenvalues are the $N$ first positive eigenvalues of $H^c+V$,
counted with multiplicity. Then $E^c_0$ is the sum of these $N$
first positive eigenvalues. Clearly,
$(W_{min},\chi_{[0,\infty)}(H^c+V))$ realizes $e^c_0$.
 \hfill $\Box$ \medskip

 The interesting case is, of course,  $\, \kappa>0\,$, when  electronic interaction is taken into account. For $\,\kappa>0\,$ and small two very different situations occur, depending on the number $\,N\,$ of electrons.

 \medskip
The  {\bf first situation}  (perturbation from the linear closed shell atom) corresponds to
\begin{equation}\label{CS} N= \Sum^I_{i = 1} N^c_i,
\quad I \in \Z^+
\end{equation}is treated in detail in \S 3.

\medskip We recall that $N^c_i$ is the dimension of the
eigenspace ${\cal M}^c_i = \mathrm{Ker} (H^c + V - \lambda_i^c
\,\1)$ already defined. Under assumption  (\ref{CS}),   for ${\kappa} = 0$, there
is a unique solution, $W^c_0$, to the variational problems
defining $E^c_0$ and $e^c_0$, $$ W^c_0 = \bigoplus^I_{i = 1} {\cal
M}^c_i . $$ The "shells" of energy $\lambda^c_{i}\;,\; 1\leq i\leq
I\;,$ are ``closed": each one is occupied by the maximal number of
electrons allowed by the Pauli exclusion principle. The subspace $
W^c_0$ is invariant under the representation $\bullet$ of $SU(2)$.

\medskip
We are interested in solutions $W^c_{\kappa}$ of the Dirac-Fock
equations lying in a neighborhood $\,\Omega\subset G_N (H^{1/2})$
of $W^c_0$, for $\kappa$ small. Using the implicit function
theorem, we are going to show that for each $\kappa$ small,
$W^c_\kappa$ exists, is unique, and is a smooth function of
$\kappa$.

Information about the properties enjoyed by $W^c_{\kappa}$ is given by

\begin{proposition}\label{[7]}
Fix $\,c\,$ large enough. Under assumption (\ref{CS}), for ${\kappa}$ small enough,
\begin{equation} \label{8}
   E^c_{\kappa} = {\cal E}_{\kappa}^c (W^c_{\kappa})
   = \inf_{W \in G_N (P^+_{\kappa,W^c_{\kappa}}H^{1/2})}
   {\cal E}_{\kappa}^c (W) ,
\end{equation}
and $W^c_{\kappa}$ is the unique solution of this minimization problem.
\end{proposition}

This proposition will be proved in \S 3. Our first main result follows from it :

\begin{theorem}\label{[8]}
Under  assumption (\ref{CS}), for $\, c>0\,
$ fixed and $\,{\kappa}\,$ small enough, $E^c_{\kappa} =
e^c_{\kappa}$ and both variational problems are achieved by the
same solution $W^c_\kappa$ of the self-consistent Dirac-Fock
equations. For $e^c_{\kappa}$, the optimal projector in
$\widetilde{\cal P}_\kappa$ is $P^{+}_{{\kappa}, W^c_{\kappa}}$.
\end{theorem}

\noindent {\bf  Proof}. The above proposition implies that for
${\kappa}$ small,
\begin{equation} \label{10}
  e^c_{\kappa} \geq  \inf_{W \in G_N (P^{+}_{{\kappa}
  , W^c_{\kappa}}H^{1/2})} {\cal E}_{\kappa}^c (W)
  ={\cal E}_{\kappa}^c(W^c_{\kappa})= E^c_{\kappa} .
\end{equation}
Therefore, $e^c_{\kappa} = E^c_{\kappa}$. Moreover, by Proposition
\ref{[7]}, $e^c_{\kappa}$ is achieved by a couple $(W^c_{\kappa},
P)$ such that $P = P^{+}_{{\kappa}, W^c_{\kappa}}$, $W^c_{\kappa}$
being a solution of the Dirac-Fock equations. This ends the proof.
$\Box$

\medskip
The {\bf second situation} (perturbation from the linear open shell case) occurs when

\begin{equation} \label{OS} N= \Sum^I_{i = 1}
N^c_i+k, \; I \in \Z^+, \; 0 < k < N^c_{I+1}\,.
\end{equation}
It is treated in detail in \S 4.

\medskip
When (\ref{OS}) holds and when
${\kappa} = 0$, there exists a manifold of solutions,  $S_0$,
whose elements are the spaces $$ \bigoplus^I_{i = 1} {\cal M}^c_i
\oplus W^c_{I+1, k} , $$ for all $W^c_{I+1, k} \in G_\ell
(M^c_{I+1})$. These spaces are all the solutions of the
variational problems defining $E^c_0$ and $e^c_0$. The $(I+1)$-th
"shell" of energy $\lambda^c_{I+1}$ is ``open": it is occupied by
$k$ electrons, while the Pauli exclusion principle would allow
$N^c_{I+1}-k$ more. {\sl Note that we use the expression ``open shell" in
the linear case $\,\kappa=0\,$  only :  } indeed,  adapting an idea of Bach {\it et
al.} \cite{BLLS}, one can easily see that for $\kappa$ {\it
positive} and small, the solutions to (DF) at the minimal level
$E^c_\kappa$ have no unfilled shells.
\medskip

For ${\kappa} > 0 $ and small we look for solutions of the DF
equations near $S_0$ (see \S 4). We could simply quote the existence results
of \cite{Esteban-Sere-02}, and show the convergence of solutions
of (DF) at level $E^c_\kappa$, towards points of $S_0$, as
$\kappa$ goes to $0$. But we prefer to give another existence
proof, using  tools from bifurcation theory.
This approach gives a more precise picture of the set of
solutions to (DF) near the level $E^c_\kappa$ (Theorem \ref{[9]}).

\medskip
In particular, we obtain in this way all the solutions of (DF) with smallest energy $\,E^c_\kappa$ (Proposition \ref{[7bis]}).

\medskip
We now choose one of these minimizers, and we call it
$W^c_{\kappa}$. We have $P^{-}_{{\kappa}, W^c_{\kappa}}
(W^c_{\kappa})=0\,.$ Since $V$ is radial, $W^c_{\kappa}$ belongs
to an $SU(2)$ orbit of minimizers. We are interested in cases
where this orbit is not reduced to a point. Then the mean-field
operator $H^c_{{\kappa}, W^c_{\kappa}}$ should not commute with
the action $\bullet$ of $SU(2)$, and one expects the following
property to hold:
\medskip

{\it {\bf (P)} : Given $c$ large enough, if $\kappa$ is small, then
for any solution $W_\kappa^c$ of (DF) at level $\,E^c_\kappa$, there is a matrix $A \in SU(2)$ such that
 \begin{equation}\label{symm}
     P^{-}_{{\kappa}, W^c_{\kappa}} (A\bullet W^c_{\kappa})\neq 0\,.
 \end{equation}
}

Let us explain why {\bf (P)} contradicts Conjecture M:

\begin{proposition} \label{[10]}
If {\bf (P)} is satisfied, then for $c$ large enough and $\kappa$
small, given any solution $W_\kappa^c$ of the nonlinear Dirac-Fock
equations such that ${\cal E}_{\kappa}^c (W^c_{\kappa}) =
E_{\kappa}$, we have
\begin{equation} \label{13}
  E^c_{\kappa} = {\cal E}^c_\kappa (W^c_{\kappa}) > \inf_{W \in G_N
  (H^{1/2}) \atop P^{-}_{{\kappa}, W^c_{\kappa}} W = 0} {\cal
  E}_{\kappa}^c (W) .
\end{equation}
\end{proposition}

This proposition will be proved in \S 4. Moreover, we verify (see Proposition \ref{conj}) that (P) holds when $\,I\geq 1\,$ and $\,k=1$, {\sl i.e.} when in the linear case there is a single electron in the highest nonempty shell.

Our second main result follows directly from Propositions \ref{[10]} and \ref{conj}.

\begin{theorem}\label{TT} Take
$$ N= \Sum^I_{i = 1} N^c_i +1,
\quad I \geq 1\,.
$$
For $\, c\,$ large and $\,\kappa>0\,$ small,  there is no solution $W_*$ of the nonlinear Dirac-Fock
equations with positive Lagrange multipliers, such that the couple
 $$
  ( W_*,P^{+}_{{\kappa}, W_*})
 $$
realizes the max-min $e^c_{\kappa}$. So Conjecture $M$ is wrong.
\end{theorem}

\vspace{6mm}

\noindent {\bf 3 - Perturbation from the linear closed shells case}.

\vspace{2mm} Let us recall that we are in the case
$$ N= \Sum^I_{i = 1} N^c_i,
\quad I \in \Z^+\,,
$$ $N^c_i$ being the dimension of the
eigenspace ${\cal M}^c_i = \mathrm{Ker} (H^c + V - \lambda_i^c
\,\1)$.
We want to apply  the implicit function
theorem in a neighborhood of $\,W^c_0\,$, for $\,\kappa\, $ small.
For this purpose, we need a local chart near $W^c_0$.
Take an orthonormal basis $(\psi_1,\cdots,\psi_N)$ of $W^c_0$,
whose elements are eigenvectors of $H^c+V$, the associated
eigenvalues being $\mu_1\leq\cdots\leq \mu_N$ ({\sl i.e.}  $\lambda_1^c, \dots, \lambda^c_I$ counted with multiplicity). Let ${\cal Z}$ be the orthogonal space of
$W^c_0$ for the $L^2$ scalar product, in
$H^{1/2}(\BbR^3,{\BbC}^4)$. Then ${\cal Z}$ is a Hilbert space for
the $H^{1/2}$ scalar product. The map
 $$
   C\,:\,\chi=(\chi_1,\cdots,\chi_N)\to
   span(\psi_1+\chi_1,\cdots,\psi_N+\chi_N)\;,
 $$
defined on a small neighborhood ${\cal O}$ of $0$ in ${\cal Z}^N$,
is the desired local chart. Denote $G_\chi$ the $N\times N$ matrix
of scalar products $(\chi_l,\chi_\ell)_{L^2}$. Then
 $$
   {\cal E}_\kappa^c\circ C(\chi)={\cal E}_\kappa^c
   \bigl((I+G_\chi)^{-1/2}(\psi+\chi)\bigr)\;.
 $$
The differential of this functional defines a smooth map
$F_\kappa\,:\,{\cal O}\subset {\cal Z}^N\to ({\cal Z}')^N$, where
${\cal Z}' \subset H^{-1/2}$ is the topological dual of ${\cal Z}$
for the $H^{1/2}$ topology, identified with the orthogonal space
of $W^c_0$ for the duality product in $H^{-1/2}\times H^{1/2}$.
Note that $F_\kappa$ depends smoothly on the parameter $\kappa$. A
subspace $C(\chi)$ is solution of (DF) if and only if
$F_\kappa(\chi)=0$. To apply the implicit function theorem, we
just have to check that the operator $L:=D_\chi F_0(0)$ is an
isomorphism from ${\cal Z}^N$ to its dual $({\cal Z}')^N$. This
operator is simply the Hessian of the DF energy expressed in our
local coordinates:
\begin{equation} \label{hessian}
   L\chi=\bigl(
  (H_c+V-\mu_1)\chi_1,\cdots,(H_c+V-\mu_N)\chi_N\bigr)\;.
\end{equation}
Under  assumption (\ref{CS}), the scalars $\mu_k,\, k=1, \dots N$, are not
eigenvalues of the restriction of $H^c+V$ to the $L^2$-orthogonal
subspace of $W^c_0$. This implies that $L$ is an isomorphism. As a
consequence, there exists a neighborhood of $W^c_0 \times \{0\}$
in $G_N (H^{1/2}) \times \BbR$, $\Omega \times (-{\kappa}_0,
{\kappa}_0)\,$ and  a smooth function $h^c : (-{\kappa}_0,
{\kappa}_0) \rightarrow \Omega$ such that for ${\kappa} \in
(-\kappa_0, {\kappa}_0)$, $\,W^c_{\kappa}:=h^c(\kappa)$ is the
unique solution of the Dirac-Fock equations in $\Omega$. Moreover,
for all $\,{\kappa}\in  (-\kappa_0,{\kappa}_0)$, the following
holds:
\begin{equation} \label{inv}
    u (W^c_{\kappa}) =
    W^c_{\kappa}\;,\;\forall u \in SU(2)\;.
\end{equation}
Indeed, the subset $A$ of parameters $\kappa$ such that
(\ref{inv}) holds is obviously nonempty (it contains $0$) and
closed in $(-\kappa_0,{\kappa}_0)$. Now, for $\kappa$ in a small
neighborhood of $A$, the $SU(2)$ orbit of $W^c_\kappa$ stays in
$\Omega$. But this orbit consists of solutions of the Dirac-Fock
equations, so, by uniqueness in $\Omega$, it is reduced to a
point. This shows that $A$ is also open. $A$ is thus the whole
interval of parameters $(-\kappa_0,{\kappa}_0)$.
\medskip

Now we are in the position to prove Proposition \ref{[7]}.

\medskip
\noindent {\bf  Proof of Proposition \ref{[7]}}.

\medskip Remember that for ${\kappa} = 0$,
$P^+_{0,W^c_{0}}$ coincides with $\chi_{(0,\infty)}(H^c+V)$. Now,
$W^c_0$ is clearly the unique minimizer of ${\cal E}^c_0$ on the
Grassmannian submanifold $G^+_0:=G_N(P^+_{0,W^c_{0}}H^{1/2})$.
More precisely, in topological terms, for any neighborhood ${\cal
V}$ of $W^c_0$ in $G_N (H^{1/2})$, there is a constant
$\delta=\delta({\cal V})>0$ such that
\begin{equation} \label{top}
  {\cal E}^c_0(W)\geq {\cal E}^c_0(W^c_0)+\delta\;,
  \;\forall \,W \in G^+_0\cap(G_N (H^{1/2})\setminus
  {\cal V})\;.
\end{equation}
Moreover, looking at formula (\ref{hessian}), one easily sees that
the Hessian of ${\cal E}^c_0$ on $G^+_0$ is positive definite at
$\, W^c_0\,$. We now take $\kappa>0$ small, and we consider again
the chart $C$ constructed above. We define the submanifold
$G^+_\kappa:=G_N(P^{ +}_{{\kappa}, W^c_{\kappa}}H^{1/2})$. Then
the restriction $C^+_\kappa$ of $C$ to $(P^{ +}_{{\kappa},
W^c_{\kappa}}{\cal Z})^N$ is a local chart of $G^+_\kappa$ near
$W^c_{\kappa}$. For ${\kappa}$ small enough, there is a
neighborhood ${\cal U}$ of $0$ in ${\cal Z}^N$ such that the
second derivative of ${\cal E}^c_{\kappa}\circ C^+_\kappa$ is
positive definite on ${\cal U}^+_\kappa:={\cal U}\cap (P^{
+}_{{\kappa}, W^c_{\kappa}}{\cal Z})^N$. The functional ${\cal
E}^c_{\kappa}\circ C^+_\kappa$ is thus strictly convex on ${\cal
U}^+_\kappa$. Now, for $\kappa$ small, there is a unique
$\chi_\kappa\in {\cal U}^+_\kappa$ such that
$C^+_\kappa(\chi_\kappa)=W^c_\kappa$. Then the derivative of
${\cal E}^c_{\kappa}\circ C^+_\kappa$ vanishes at $\chi_\kappa$.
As a consequence $W^c_{\kappa}=C^+_\kappa(\chi_\kappa)$ is the
unique minimizer of ${\cal E}^c_{\kappa}$ on ${\cal
V}^+_\kappa:=C^+_\kappa({\cal U}^+_\kappa)$. Now, we choose, as
neighborhood of $W^c_0$ in $G_N (H^{1/2})$, the set ${\cal
V}:=C({\cal U})$, and we consider the constant $\delta>0$ such
that (\ref{top}) is satisfied. Taking $\kappa>0$ even smaller, we
can impose
 $$
   \Min_{{\cal V}^+_\kappa} {\cal E}_{\kappa}^c
   +\delta/2 \leq  \Inf _{G^+_\kappa\setminus {\cal V}^+_\kappa}
   {\cal E}^c_{\kappa} .
 $$
Hence, $W^c_\kappa$ is the unique solution to the minimization
problem (\ref{8}).
 $\Box$
 \vspace{6mm}

\noindent {\bf 4 - Bifurcation from the linear open shell case}.

\vspace{2mm} Recall that here we are in the case
$$ N= \Sum^I_{i = 1}
N^c_i+k, \; I \in \Z^+, \; 0 < k < N^c_{I+1}\,.
$$
For
${\kappa} = 0$, there exists a manifold of solutions,  $S_0$,
whose elements are the spaces $$ \bigoplus^I_{i = 1} {\cal M}^c_i
\oplus W^c_{I+1, k} , $$ for all $W^c_{I+1, k} \in G_\ell
(M^c_{I+1})$. These spaces are all the solutions of the
variational problems defining $E^c_0$ and $e^c_0$.

\medskip
For ${\kappa} > 0 $ and small we want to find solutions of the DF
equations near $S_0$,  by using  tools from bifurcation theory.

If $\lambda_{I+1}$ has only multiplicity 2, then (\ref{OS}) implies $k=1$
and by Lemma 1
of \S 2, $S_0$ is an $SU(2)$ orbit. Then, as in \S 3, one can
find, in a neighborhood of $S_0$, a unique $SU(2)$ orbit
$S_\kappa$ of solutions of (DF). But there are also more degenerate
cases in which $\lambda_{I+1}$ has a higher multiplicity, and $S_0$
contains a continuum of $SU(2)$ orbits. In such situations,
$\kappa=0$ is a bifurcation point, and one expects, according to
bifurcation theory, that the manifold of solutions $S_0$ will
break up for $\kappa\neq 0$, and that there will only remain a
finite number of $SU(2)$ orbits of solutions. To find these
orbits, one usually starts with a  Lyapunov-Schmidt reduction: one
builds a suitable manifold $S_\kappa$ which is diffeomorphic to
$S_0$ (see {\it e.g.} \cite{Ambros}). When $S_0$ contains several
$SU(2)$ orbits, the points of $S_\kappa$ are not necessarily
solutions of (DF), but $S_\kappa$ contains all the solutions
sufficiently close to $S_0$. Moreover, all critical points of the
restriction of ${\cal E}^c_\kappa$ to $S_\kappa$ are solutions of
(DF). The submanifold $S_\kappa$ is constructed thanks to the
implicit function theorem. More precisely, we consider the
projector $\Pi\,:\;L^2\to \bigoplus^{I+1}_{i = 1} {\cal M}^c_i$.
To each point $z\in S_0$ we associate the submanifold $F_z:=\{w\in
G_N(H^{1/2})\;:\;\Pi w=z\}$. For $w$ a point of $F_z$, let
$\Delta_w:=T_w F_z\subset T_w G_N(H^{1/2})$. Then the following
holds:

\begin{theorem}\label{[9]} Under the above assumptions,
there exist a neighborhood $\Omega$ of $S_0$ in $G_N(H^{1/2})$, a
small constant $\kappa_0>0$, and a smooth function
$h\,:\;S_0\times (-\kappa_0,\kappa_0)\to\Omega$ such that
\begin{itemize}
\item[(a)] $h (z,0) = z \quad \forall z \in S_0$
\item[(b)] Denoting $S_\kappa:=h(S_0,\kappa)$,
$S_\kappa$ is also the set of all points $w$ in $\Omega$ such that
\begin{equation}\label{half}
  <({\cal E}^c_\kappa)'(w), \xi>=0, \quad \forall
  \xi \in \Delta_w
\end{equation}
\item[(c)] $ h(z,\kappa)\in F_z\;, \quad \forall (z,
\kappa) \in S_0 \times (- \kappa_0, \kappa _0)$.
\end{itemize}
\end{theorem}
\noindent {\bf  Proof}.

We first fix a point $z$ in $S_0$. Let ${\cal N}$ be the
orthogonal space of $\bigoplus^{I+1}_{i = 1} {\cal M}^c_i$ in
$H^{1/2}$ for the $L^2$ scalar product. As in \S 3, we can define
a local chart $C_z\,:\;{\cal O}\subset ({\cal N})^N\to F_z$ near
$z$, by the formula $C(\chi)=span(\psi+\chi)$, where
$\psi=(\psi_1,\cdots,\psi_N)$ is an orthonormal basis of $z$
consisting of eigenvectors of $H^c+V$, with eigenvalues
$\mu_1\leq\cdots\leq\mu_N$ ({\sl i.e.}  $\lambda_1^c, \dots, \lambda^c_I$ counted with multiplicity). The Hessian of  ${\cal E}^c_0\circ
C_z$ at $\chi=0$ is given once again by formula (\ref{hessian}).
It is an isomorphism between $({\cal N})^N$ and its dual. So,
arguing as in \S 3, we find, by the implicit function theorem, a
small constant $\kappa_z>0$, a neighborhood $ \omega_z$ of $z$ in
$F_z$ and a function $\tilde{h}_z\,:\,(-\kappa_z,\kappa_z)\to
\widetilde{\Omega}_z$ such that:
\begin{itemize}
\item[(i)] $\tilde{h}_z (0) = z$
\item[(ii)] $\tilde{h}_z(\kappa)$  is the unique point
$w$ in $ \widetilde{\Omega}_z$ such that
\begin{equation}
    <({\cal E}^c_\kappa)'(w), \xi>=0, \quad \forall
    \xi \in \Delta_w
\end{equation}
\end{itemize}
Since $S_0$ is compact and ${\cal E}^c_\kappa(w)$ a smooth
function of $(w,\kappa)$, it is possible to choose $\kappa_z\;,\
\widetilde{\Omega}_z$ such that $\kappa_0:=\inf_{z\in S_0}\kappa_z
>0$, with $\Omega:=\bigcup_{z\in S_0} \widetilde{\Omega}_z$ a
neighborhood of $S_0$,  and $h(z,\kappa):= \tilde{h}_z(\kappa)$ a
smooth function on
 $S_0\times (-\kappa_0,\kappa_0)$ with values in $\Omega$. This function satisfies (a,b,c).
$\Box$
\medskip

From (b) any critical point of  $ {\cal E}^c_\kappa$ in $\Omega$
must lie on $S_\kappa$. From (c) it follows that $S_\kappa$ is a
submanifold diffeomorphic to $S_0$, and transverse to each fiber
$F_z$ in $G_N(H^{1/2})$. If $z\in S_0$ is a critical point of
${\cal E}^c_\kappa\circ h(\cdot,\kappa)$, then, taking
$w=h(z,\kappa)$, the derivative of ${\cal E}^c_\kappa$ at $w$
vanishes on $T_w S_\kappa$. From (b), it also vanishes on the
subspace $\Delta_w$ which is transverse to $T_w S_\kappa$ in $T_z
G_N(H^{1/2})$, hence $({\cal E}^c_\kappa)'(w)=0$. This shows that
the set of critical points of ${\cal E}^c_\kappa$ in $\Omega$
coincides with the set of critical points of the restriction of
${\cal E}^c_\kappa$ to $S_\kappa$. Arguing as in the proof of
Proposition \ref{[7]}, one gets more:

\begin{proposition}\label{[7bis]}
For ${\kappa} > 0$ small, the solutions of (DF) of smallest energy
$E^c_\kappa$ are exactly the minimizers of ${\cal E}^c_\kappa$ on
$S_{\kappa}$.\end{proposition}

We are now ready to prove Proposition \ref{[10]}.

\medskip
\noindent {\bf  Proof of Proposition \ref{[10]}
}. Since ${\kappa}$ is small, for any matrix
$A\in SU(2)$ the map $P^+_{{\kappa}, A\bullet W^c_{\kappa}}$
induces a diffeomorphism between the submanifolds $G_N
(P^{+}_{{\kappa}, W^c_{\kappa}}H^{1/2})$ and $G_N
(P^{+}_{{\kappa},\, A\bullet W^c_{\kappa}}H^{1/2})\,$.

Now, we fix $A \in SU(2)$ such that (\ref{symm}) holds. Then there
exists a unique point $W^+ \in G_N (H^{1/2})$ such that
\begin{equation} \label{15}
  P^{-}_{{\kappa}, W^c_{\kappa}} W^+ = 0 , \quad P^{+}_{{\kappa}, \,A\bullet
  W^c_{\kappa}} W^+ = A\bullet W^c_{\kappa}
\end{equation}
By (\ref{symm}), we have
 $$
  W^+ \not = A\bullet W^c_{\kappa} .
 $$
On the other hand, in \cite{Esteban-Sere-01} it was proved that
\begin{equation}\label{16}
   {\cal E}_{\kappa}^c (A\bullet W^c_{\kappa}) = \sup_{W \in G_N
   (H^{1/2}) \atop P^+_{{\kappa}, A\bullet W^c_{\kappa}} W = A\bullet
   W^c_{\kappa}} {\cal E}_{\kappa}^c (W)
\end{equation}
and $A\bullet W^c_{\kappa}$ is the unique solution of this
maximization problem. Therefore,
 $$
   {\cal E}_{\kappa}^c (A\bullet W^c_{\kappa}) > {\cal
   E}_{\kappa}^c (W^+)\;.
 $$
But
 $$
  {\cal E}^c_{\kappa} (W^+) \geq \inf_{W \in  G_N
  (P^{+}_{{\kappa}, W^c_{\kappa}}H^{1/2})} {\cal E}_{\kappa}^c
  (W)\;,
 $$
hence, by invariance of ${\cal E}_{\kappa}^c$ under the action of
$SU(2)$,
 $$
  E^c_{\kappa}  = {\cal E}_{\kappa}^c (A\bullet
  W^c_{\kappa}) > \inf_{W \in  G_N (P^{+}_{{\kappa},
  W^c_{\kappa}}H^{1/2})} {\cal E}_{\kappa}^c (W)\; ,
 $$
and the Proposition is proved. \hfill $\Box$

\medskip

Since there are no solutions of (DF) under level $E^c_\kappa$, and
$e^c_\kappa\leq E^c_\kappa$, Proposition \ref{[10]} has the
following consequence:

\begin{corollary}\label{last}
If {\bf (P)} is satisfied, then for $c$ large enough and $\kappa$
small, there is no solution $W_*$ of the nonlinear Dirac-Fock
equations with positive Lagrange multipliers, such that the couple
 $$
  ( W_*,P^{+}_{{\kappa}, W_*})
 $$
realizes the max-min $e^c_{\kappa}$. So Conjecture $M$ is wrong when {\bf (P)} holds.
\end{corollary}
\medskip
\bigskip

We now exhibit a case where {\bf (P)} holds.

\begin{proposition}\label{conj}
Assume that $\,N= \Sum^I_{i = 1} N^c_i+1,\; I \geq 1.$ Then {\bf (P)} is satisfied.
\end{proposition}

\medskip

\noindent {\bf  Proof}.
 \medskip

\noindent {\bf  Step 0}. Fix $\,c\,$ large enough and take a
sequence of positive parameters $(\kappa_\ell)_{\ell\geq 0}$ converging
to $0$. Let $(W^c_\ell)_{\ell\geq 0}$ be a sequence in $G_N(H^{1/2})$,
with $W^c_\ell$ a minimizer of ${\cal E}^c_{\kappa_\ell}$ on
$S_{\kappa_\ell}$. Let $\, \psi_\ell^c\in W_{\ell}^c$ be an eigenvector of
the mean-field Hamiltonian $H^c_{\kappa_\ell,W^c_\ell} $, normalised in
${L^2}$ and corresponding to the highest occupied level.
Extracting a subsequence if necessary, we may assume that $\,
\psi^c_{\ell} \to \psi^c \in {\cal
M}^c_{I+1}=\mathrm{Ker}(H^c+V-\lambda^c_{I+1})$.  Moreover, from
Theorem \ref{[9]} we have
 $$
   W_{\ell}^c\to W^c_0=\bigoplus^I_{i = 1}
   {\cal M}^c_i \oplus \BbC\; \psi^c \; .
 $$

\noindent {\bf  Step 1}. Fix $\, c\geq 1\,$. Since
$P^{-}_{{\kappa_\ell},W^c_{\ell}} \psi^c_{\ell}=0$, we can write, by a
classical result due to Kato,
\begin{equation}\label{proj}
  \quad P^{-}_{{\kappa_\ell},A\bullet W^c_{\ell}}
  \psi^c_{\ell}=\frac{1}{2\pi}\int_{-\infty}^{+\infty}\!\!\left[
  (H^c_{{\kappa_\ell},W^c_{\ell}}-i\eta)^{-1}\!\!-
  \!(H^c_{{\kappa_\ell},A\bullet W^c_{\ell}}-i\eta)^{-1}
  \right]\psi^c_{\ell}\,d\eta
\end{equation}
 $$
  \quad =\frac{1}{2\pi}\!\int_{-\infty}^{+\infty}
  \!(H^c_{{\kappa_\ell},W^c_\ell}-i\eta)^{-1}(H^c_{{\kappa_\ell},A\bullet
  W^c_{\ell}}\!-\!H^c_{{\kappa_\ell},W^c_\ell}) (H^c_{{\kappa_\ell},A\bullet
  W^c_\ell}-i\eta)^{-1}  \psi^c_\ell\,d\eta
 $$
 $$
  \qquad =
  \frac{\kappa_\ell}{2\pi}\int_{-\infty}^{+\infty}\!(H^c+V-i\eta)^{-1}(\Omega_{A\bullet
  W^c_0}-\Omega_{W^c_0}) (H^c+V-i\eta)^{-1}  \psi^c \;d\eta
  +o({\kappa_\ell}) \,,
 $$
where by $\,\Omega_{W}\,$ we denote the nonlinear part of
$\,H^c_{{\kappa}, W}$:
 $$
  \,H^c_{{\kappa}, W}=H^c+V+\kappa\,
  \Omega_W\,.
 $$

But  note that since the space $\,\bigoplus^I_{i = 1} {\cal M}^c_i
\,$ is invariant under the action of $\,SU(2)$,
 $$
   \Omega_{A\bullet
   W^c_0}-\Omega_{W^c_0}=\Omega_{A\bullet
   \psi^c}-\Omega_{\psi^c}\,.
 $$

So, we just have to prove that for $\, c\,$ sufficiently large and
for all $\,\psi^c\in M^c_{I+1}\,$, there exists $\,A\in SU(2)\,$
such that
\begin{equation}\label{adem}
  \int_{-\infty}^{+\infty}\!(H^c+V-i\eta)^{-1}(\Omega_{A\bullet
  \psi^c}-\Omega_{\psi^c}) (H^c+V-i\eta)^{-1}  \psi^c \;d\eta \ne
  0\,.
\end{equation}
Since
 $$
   (H^c+V-i\eta)^{-1}\psi^c=\frac {
   \psi^c}{\lambda^c_{I+1}-i\eta}\quad \mbox{and}\quad
   \Omega_{\psi^c}\,\psi^c=0\,,
 $$
what we need to prove is that for all nonzero $\,\psi^c\in
M^c_{I+1}\,$,  there exists $\,A\in SU(2)\,$ such that $\, {\cal
L}^c(\Omega_{A\bullet \psi^c}\,\psi^c)\ne 0$, with
 $$
   {\cal L}^c:=\int_{-\infty}^{+\infty}
   (H^c+V-i\eta)^{-1}
   \frac{d\eta}{\lambda^c_{I+1}-i\eta}\,.
 $$

\medskip

\noindent {\bf  Step 2}. We give an asymptotic expression for $\,
{\cal L}^c\,$ when $\,c\to +\infty$:
\begin{equation}
   {\cal L}^c=\frac 1{c^2}\int_{-\infty}^{+\infty}\Big(\frac
   {1}{c^2}(H^c+V)-i\frac{\eta}{c^2} \Big)^{-1}\frac{d(\eta/c^2)}
   {\frac{\lambda^c_{I+1}}{c^2}-i\frac{\eta}{c^2}}=\frac 1{c^2}\left(
   L_c +O\Big(\frac 1{c^2}\Big)\right)\,,
\end{equation}
where $L_c$, in the Fourier domain, is the operator of
multiplication by the matrix
\begin{equation}
  \hat{L}_c(p)=\int_{-\infty}^{+\infty} (-iu+\beta+
  (\boldsymbol{\alpha}\cdot p)/c)^{-1}(-iu+1)^{-1}\,du\,.
\end{equation}
Here, we have used the standard fact that
 $$
   \frac{\lambda^c_{I+1}}{c^2}= 1+O\Big(\,\frac{1}{c^2}\,\Big)\,.
 $$
\noindent
We have
 $$
   (-iu+\beta+(\boldsymbol{\alpha}\cdot p)/c)^{-1}
   =\frac{1}{-iu+\omega^c(p)} \hat{\Lambda}^c_+(p)
   + \frac{1}{-iu-\omega^c(p)} \hat{\Lambda}^c_-(p)
 $$
with
 $$
   \omega^c(p):=\sqrt{1+|p|^2/c^2}\;,
   \quad \hat{\Lambda}^c_\pm(p)=\frac{\omega^c(p)
   \pm(\beta+(\boldsymbol{\alpha}\cdot p)/c)}{2\omega^c(p)}\;.
 $$
\noindent
Hence, by the residues theorem,
 $$
   \frac{2}{\pi}\hat{L}_c(p)=\beta-1+\frac{(\boldsymbol{\alpha}
   \cdot p)}{c}
   +O\Big(\,\frac{|p|^2}{c^2}\,\Big)\,.
 $$
 \medskip

\noindent {\bf  Step 3}. It is well known (see \cite{Thaller-92})
that $\,\psi^c$ can be written as
 $$
   \psi^c={{ \phi}\choose
   {\frac{-i(\sigma\cdot \nabla)\phi}{2c} }}+
   O\Big(\,\frac{1}{c^2}\,\Big)\,,
 $$
$\phi\in L^2(\BbR^3, \BbC^2)$ being an eigenstate of
$(\frac{-\Delta}{\;2}+V)$, with eigenvalue $\, \mu=\lim_{c\to
+\infty} \,(\lambda^c_{I+1}-c^2).$ Since we have assumed that $V$
is smooth, this asymptotic result holds for the topology of the
Schwartz space ${\cal S}(\BbR^3)$. So,
 $$
   \frac{2c^2}{\pi}{\cal
   L}^c(\Omega_{A\bullet\psi^c}\,\psi^c)
   =\frac {i}{c}{Ê{0} \chooseÊ{
   f(A,\phi)  }}+
   O\Big(\frac{1}{c^2}\Big)\,,
 $$
where
\begin{equation}
  \quad f(A,\phi):=\Big(|A\bullet\phi|^2*\frac
  {x\cdot\sigma}{|x|^3}\Big)\phi -
  \Big(\!\!<A\bullet\phi,\phi>_{\BbC^2}*\frac
  {x\cdot\sigma}{|x|^3}\Big)(A\bullet\phi)\;.
\end{equation}

\noindent
What remains to prove is :

 \medskip

\noindent{\bf Step 4}.  For any eigenvector $\,\phi$ of  the
Schr\"odinger operator $-\frac{\Delta}{2}+V$, there exists an
$\,A\in SU(2)\,$ such that $f(A,\phi) \not \equiv 0\,.$
\bigskip

\noindent {\bf  Proof of Step 4}. We consider the integral $$
I_{A,\phi}(r):=\int_{S^2} <(x\cdot
\sigma)\phi,f(A,\phi)>_{\BbC^2}(r\, \omega)d\omega\;.$$ Since
$\phi$ has exponential fall-off at infinity, the electrostatic
field $|A\bullet\phi|^2*\frac {x}{|x|^3}$ takes the asymptotic
form $\Big(\int_{\BbR^3}|A\bullet\phi|^2\Big)\frac
{x}{|x|^3}+O\Big(\frac{1}{|x|^3}\Big)$ when $|x|$ is large. The
same phenomenon holds for the convolution product
$<A\bullet\phi,\phi>_{\BbC^2}*\frac {x}{|x|^3}$ . As a
consequence, for $r$ large,

\begin{eqnarray*}
   r\, I_{A,\phi}(r)&= & \Big(\int_{\BbR^3}
   |A\bullet\phi|^2\Big)\Big(\int_{S^2}|\phi |^2
   (r\,\omega)\,d\omega\Big )\\
   & & - \Big(\int_{\BbR^3}<A\bullet\phi,\phi>_{\BbC^2}\Big)
   \Big(\int_{S^2}<\phi,A\bullet\phi>_{\BbC^2}(r\,\omega)
   \,d\omega\Big)\\
   &&+O\Big(\frac{1}{r}\Big)\Big(\int_{S^2}|\phi|^2(r\,\omega)
   \,d\omega\Big )\;.
\end{eqnarray*}
Since $\bullet$ is unitary, the Cauchy-Schwartz inequality gives
 $$
   \int_{S^2}|\phi |^2(r\,\omega)\,d\omega=
   \int_{S^2}|A\bullet \phi
   |^2(r\,\omega)\,d\omega\geq \Big |
   \int_{S^2}<A\bullet\phi,\phi>_{\BbC^2}(r\,\omega)\,d\omega \Big |
    \;.
 $$
By Lemma 1 of \S1, we can choose $A$ such that $\phi$ and
$A\bullet \phi$ are not colinear. Then
$$\int_{\BbR^3}|A\bullet\phi|^2\;=\;\int_{\BbR^3}|\phi|^2
\quad>\quad\Big | \int_{\BbR^3}<A\bullet\phi,\phi>_{\BbC^2} \Big |
\;.$$ So there is a constant $\delta > 0$ such that, for $r$ large
enough,
\begin{equation}\label{delta}
  | r\, I_{A,\phi}(r) |\geq
  \delta \Big(\int_{\BbR^3}|\phi|^2\Big) \Big(\int_{S^2}|\phi
  |^2(r\,\omega)\,d\omega\Big )\;.
\end{equation}

Being an eigenvector of the Schr\"odinger operator
$-\frac{\Delta}{2}+V$, the function $\phi$ cannot have compact
support. So the lower estimate (\ref{delta}) implies that the
function $I_{A,\phi}(r)$ is not identically $0$, hence $f(A,\phi)
\not\equiv 0\,.$ Step 4 is thus proved, and {\bf (P)} is
satisfied. $\Box$

\bigskip
{\bf Aknowledgement.} The authors wish to thank the referee for useful comments on the first version of this paper.

\end{document}